\documentclass[oneside,11pt]{article}

\usepackage[utf8]{inputenc}
\usepackage[T1]{fontenc}

\usepackage{amsthm}
\usepackage{amsmath}
\usepackage{amsfonts}
\usepackage{amssymb}

\usepackage{graphicx} 
\usepackage{float}
\usepackage{booktabs}
\usepackage{multirow}
\usepackage{rotating}
\usepackage{array}
\usepackage{xcolor}
\usepackage{subfig}
\usepackage[ruled]{algorithm2e}

\newcolumntype{R}[1]{>{\raggedleft\let\newline\\\arraybackslash\hspace{0pt}}m{#1}}
\newcolumntype{L}[1]{>{\raggedright\let\newline\\\arraybackslash\hspace{0pt}}m{#1}}

\usepackage{breakcites}

\usepackage{nowidow}

\usepackage{enumitem}

\usepackage[top=1.5cm,bottom=2cm,right=2.5cm,left=2.5cm]{geometry}
\usepackage{titling}

\usepackage{natbib}

\usepackage{ifplatform}

\usepackage{textcomp}

\ifwindows
\fi

\allowdisplaybreaks

\DeclareFontFamily{OT1}{pzc}{}
\DeclareFontShape{OT1}{pzc}{m}{it}{<-> s * [1.10] pzcmi7t}{}
\DeclareMathAlphabet{\mathpzc}{OT1}{pzc}{m}{it}

\usepackage[pdftex,final,bookmarksnumbered,bookmarksopen=false,breaklinks,colorlinks]{hyperref}
\hypersetup{
	final,
	bookmarksnumbered=true,
	bookmarksopen=true,
	bookmarksopenlevel=0,
	unicode=false,          
	pdftoolbar=true,        
	pdfmenubar=true,        
	pdffitwindow=false,     
	pdftitle={A review of goodness-of-fit tests for models involving functional data},    
	pdfdisplaydoctitle=true, 
	pdftoolbar=true,
	pdfmenubar=true,
	pdflang={English},
	pdfauthor={Wenceslao Gonzalez-Manteiga, Rosa M. Crujeiras, Eduardo Garcia-Portugues},     
	pdfsubject={arXiv paper},   
	pdfcreator={Eduardo Garcia-Portugues},   
	pdfproducer={Eduardo Garcia-Portugues}, 
	pdfkeywords={Distribution}{Functional data}{Goodness-of-fit}{Regression}{Test}, 
	pdfnewwindow=true,      
	breaklinks=true,
	hidelinks,       
	linkcolor=black,          
	citecolor=black,        
	filecolor=magenta,      
	urlcolor=cyan           
}

\newif\ifmain
\maintrue
\newif\ifsupplement
\supplementtrue

\newif\iffigstabs
\figstabstrue

\begin{document}

\ifmain

\title{A review of goodness-of-fit tests for models involving functional~data}
\setlength{\droptitle}{-1cm}
\predate{}%
\postdate{}%
\date{}

\author{Wenceslao Gonz\'alez-Manteiga$^{1,3}$,  Rosa M. Crujeiras$^{1}$, and Eduardo Garc\'ia-Portugu\'es$^{2}$}
\footnotetext[1]{Department of Statistics, Mathematical Analysis and Optimization, University of Santiago de Compostela (Spain).}
\footnotetext[2]{Department of Statistics, Carlos III University of Madrid (Spain).}
\footnotetext[3]{Corresponding author. e-mail: \href{mailto:wenceslao.gonzalez@usc.es}{wenceslao.gonzalez@usc.es}.}
\maketitle

\begin{abstract}
	A sizable amount of goodness-of-fit tests involving functional data have appeared in the last decade. We provide a relatively compact revision of most of these contributions, within the independent and identically distributed framework, by reviewing goodness-of-fit tests for distribution and regression models with functional predictor and either scalar or functional response.
\end{abstract}
\begin{flushleft}
\end{flushleft}

\section{Introduction}
\label{GonzalezManteiga:1}

Since the earliest Goodness-of-Fit (GoF) tests were introduced by Pearson more than a century ago, there has been a prolific statistical literature on this topic. If we were to highlight a milestone in this period, that may be 1973, with the publication of \cite{Durbin73} and \cite{Bickel73}, introducing a novel design of GoF tests based on distances between distribution and density estimates, respectively.\\

To set the context for the reader, assume that $\{X_1,\ldots,X_n\}$ is an identically and identically distributed (iid) sample of a random variable $X$ with (unknown) distribution $F$ (or density $f$, if that is the case). If the target function is the distribution $F$, then the GoF testing problem can be formulated as testing $H_0:F\in\mathcal{F}_{\Theta}=\{F_{\theta}:\theta\in\Theta\subset\mathbb{R}^q\}$ vs. $H_1:F\notin\mathcal{F}_{\Theta}$, where $\mathcal{F}_{\Theta}$ stands for a parametric family of distributions indexed in some finite-dimensional set $\Theta$. A general test statistic for this problem can be written as $T_n=T(F_n,F_{\hat\theta})$, with the functional $T$ denoting, here and henceforth, some kind of distance between a nonparametric estimate, given in this case by the empirical cumulative distribution function $F_n(x)=n^{-1}\sum_{i=1}^n\mathbb{I}(X_i\leq x)$, and an estimate obtained under the null hypothesis $H_0$, $F_{\hat\theta}$ in this case. Similarly, for testing the GoF of a certain parametric density model, the testing problem is formulated as $H_0:f\in\mathpzc{f}_\Theta=\{f_{\theta}:\theta\in\Theta\subset\mathbb{R}^q\}$ vs. $H_1:f\notin\mathpzc{f}_\Theta$ and can be approached with the general test statistic $T_n=T(f_{nh},f_{\hat\theta})$. In this setting, $f_{\hat\theta}$ is the density estimate under $H_0$ and $f_{nh}$ denotes the kernel density estimator $f_{nh}(x)=n^{-1}\sum_{i=1}^nK_h(x-X_i)$ by \cite{Parzen62} and \cite{Rosenblatt56}, where $K_h(\cdot)=K(\cdot/h)/h$, $K$ is the kernel function, and $h$ is the bandwidth.\\

The previous ideas were naturally generalized to the context of regression models in the 1990s. Consider a nonparametric, random design, regression model such that $Y=m(X)+\varepsilon$ with $(X,Y)\in\mathbb{R}^{p}\times\mathbb{R}$, and $m(x)=\mathbb{E}[Y|X=x]$ and $\mathbb{E}[\varepsilon|X=x]=0$. Denote by $\{(X_i,Y_i)\}_{i=1}^n$ an iid sample of $(X,Y)$ satisfying such a model. In this context, the GoF goal is to test $H_0: m\in\mathcal{M}_{\Theta}=\{m_{\theta}:\theta\in\Theta\subset\mathbb{R}^{q}\}$ vs. $H_1: m\notin\mathcal{M}_{\Theta}$, where $\mathcal{M}_{\Theta}$ represents a parametric family of regression functions indexed in $\Theta$. Continuing along the testing philosophies advocated by \cite{Durbin73} and \cite{Bickel73}, the seminal works of \cite{Stute97} and \cite{Hardle93} respectively introduced two types of GoF tests for regression models:
\begin{enumerate}
	\item[a)] Tests based on empirical regression processes, considering distances between estimates of the integrated regression function $I(x)=\int_{-\infty}^xm(t)\,\mathrm{d}F(t)$ ($F$ being the marginal distribution of $X$ under $H_0$ and $H_1$). Specifically, the test statistics are constructed as $T_n=T(I_{n},I_{\hat\theta})$, with $I_n(x)=n^{-1}\sum_{i=1}^n\mathbb{I}(X_i\leq x)Y_i$ and $I_{\hat\theta}(x)=n^{-1}\sum_{i=1}^n\mathbb{I}(X_i\leq x)m_{\hat\theta}(X_i)$. 
	\item[b)] Smoothing-based tests, using distances between estimated regression functions, $T_n=T(m_{nh},m_{\hat\theta})$, with $m_{nh}$ a smooth regression estimator. As a particular case, $m_{nh}(x)=\sum_{i=1}^nW_{nh,i}(x)Y_i$, with $W_{nh,i}(x)$ some weights depending on a smoothing parameter $h$. Such an estimator can be obtained with Nadaraya--Watson or local linear weights (see, e.g., \cite{Wand95}). 
\end{enumerate}

A complete review of GoF for regression models was presented by \cite{Gonzalez13}, who described the aforementioned two types of paradigms and focused on the smoothing-based alternative for discussing their properties (asymptotic behavior and calibration of the distribution in practice). The authors thoroughly checked references in the statistical literature for more than two decades, and they also identified some areas where GoF tests where still to be developed. One of these areas is functional data analysis. \\

The goal of this work is to round off this previous review, updating the more recent contributions in GoF for distribution and regression models with functional data. Consequently, the rest of the chapter is organized as follows: Section \ref{GonzalezManteiga:2} is devoted to GoF for distribution models of functional random variables, and Section \ref{GonzalezManteiga:3} focuses on regression models with scalar (Section \ref{GonzalezManteiga:31}) and functional response (Section~\ref{GonzalezManteiga:32}).

\section{GoF for distribution models for functional data}
\label{GonzalezManteiga:2}

Owing to the requirement of appropriate tools for analyzing high-frequency data, and the boost provided by the books by \cite{Ramsay05} and \cite{Ferraty06} (or, more recently, by \cite{Hor12}), functional data analysis is nowadays one of the most active research areas within statistics.\\

Actually, the pressing needs of developing new statistical tools with data in general spaces have reclaimed separable Hilbert spaces as a very natural and common framework. However, given the generality of this kind of spaces, there is a scarcity of parametric distribution models for a Hilbert-valued random variable $X$ aside the popular framework of Gaussian processes.\\

Let $\mathcal{H}$ denote a Hilbert space over $\mathbb{R}$, the norm of which is given by its scalar product as $\|x\|=\sqrt{\langle x,x\rangle}$. Consider $\{X_1,\ldots,X_n\}$ iid copies of the random variable $X:(\Omega,\mathcal{A})\rightarrow(\mathcal{H},\mathcal{B}(\mathcal{H}))$, with $(\Omega,\mathcal{A},\mathbb{P})$ the probability space where the random sample is defined and $\mathcal{B}(\mathcal{H})$ the Borel $\sigma$-field on $\mathcal{H}$. The general GoF problem for the distribution of $X$ consists on testing $H_0:\mathrm{P}_{X}\in\mathcal{P}_\Theta=\{\mathrm{P}_\theta:\theta\in\Theta\}$ vs. $H_0:\mathrm{P}_{X}\notin\mathcal{P}_\Theta$, where $\mathcal{P}_\Theta$ is a class of probability measures on $\mathcal{H}$ indexed in a parameter set $\Theta$, now possibly infinite-dimensional, and $\mathrm{P}_{X}$ is the (unknown) probability distribution of $X$ induced over $\mathcal{H}$.\\

When the goal is to test the simple null hypothesis $H_0:\mathrm{P}_{X}\in\{\mathrm{P}_0\}$, a feasible approach that enables the construction of test statistics is based on projections $\pi:\mathcal{H}\rightarrow\mathbb{R}$, in such a way that the test statistics are defined from the projected sample $\{\pi(X_1),\ldots,\pi(X_n)\}$. Such an approach can be taken on the distribution function: $T_{n,\pi}=T(F_{n,\pi},F_{0,\pi})$ with $F_{n,\pi}(x)=n^{-1}\sum_{i=1}^n\mathbb{I}(\pi(X_i)\leq x)$ and $F_{0,\pi}(x)=\mathbb{P}_{H_0}(\pi(X)\leq x)$. Some specific examples are given by the adaptation to this context of the Kolmogorov--Smirnov, Cramer--von Mises, or Anderson--Darling type tests. As an alternative, and mimicking the smoothing-based tests presented in Section \ref{GonzalezManteiga:1}, a test statistic can also be built as $T_{n,\pi}=T(f_{nh,\pi},\mathbb{E}_{H_0}[f_{nh,\pi}])$ with $f_{nh,\pi}(x)=n^{-1}\sum_{i=1}^nK_h(x-\pi(X_i))$. It should be also noted that, when embracing the projection approach, the test statistic may take into account `all' the projections within a certain space, e.g. by considering $T_n=\int T_{n,\pi}\,\mathrm{d}W(\pi)$ for $W$ a probability measure on the space of projections, or take just $T_n=T_{n,\hat\pi}$ with $\hat\pi$ being a randomly-sampled projection from a certain non-degenerate probability measure~$W$.\\

Now, when the goal is to test the composite null hypothesis $H_0:\mathrm{P}_X\in\mathcal{P}_\Theta$, the previous generic approaches are still valid if replacing $\mathrm{P}_{0,\pi}(x)$ with $\mathrm{P}_{\hat\theta,\pi}(x)=\mathbb{P}_{\mathrm{P}_{\hat\theta}}(\pi(X)\leq x)$. Within this setting, \cite{Cuesta06} and \cite{Cuesta07} provide a characterization of the composite null hypothesis by means of random projections, and provide a bootstrap procedure for calibration, as well as \cite{Bugni09}. As an alternative, \cite{Dit18} follows a finite-dimensional approximation. Note that, in the space of real square-integrable functions $\mathcal{H}=L^2[0,1]$, as a particular case one may take $\pi_h(x)=\langle x,h\rangle$, with $h\in\mathcal{H}$. The previous references provide some approaches for the calibration under the null hypothesis of the rejection region $\{T_n>c_{\alpha}\}$, where $\mathbb{P}(T_n>c_{\alpha})\leq\alpha$.\\

A relevant alternative to the procedures based on projections is the use of the so-called `energy statistics' \cite{Sek17}. Working with $\mathcal{H}$ a general Hilbert separable space (as it can be seen in \cite{Lyons13}), if $X\sim \mathrm{P}_{X}$ and $Y\sim \mathrm{P}_{Y}=\mathrm{P}_0$ ($\mathrm{P}_0$ being the distribution under the null), then
\begin{align}
	E=E(X,Y)=
	2\mathbb{E}[\|X-Y\|]-\mathbb{E}[\|X-X'\|]-\mathbb{E}[\|Y-Y'\|]\geq 0, \label{eq:energy}
\end{align}
with $\{X,X'\}$ and $\{Y,Y'\}$ iid copies of the variables with distributions $\mathrm{P}_X$ and $\mathrm{P}_Y$, respectively. Importantly, \eqref{eq:energy} equals $0$ if and only if $\mathrm{P}_X=\mathrm{P}_Y$, a characterization that serves as basis for a GoF test. The energy statistic in \eqref{eq:energy} can be empirically estimated from a sample $\{X_1,\ldots,X_n\}$ as
\begin{align*}
	\hat{E}^*=2\sum_{i=1}^n\sum_{j=1}^n\|X_i-Y_j^*\|-\sum_{i=1}^n\sum_{j=1}^n\|X_i-X_j\|-\sum_{i=1}^n\sum_{j=1}^n\|Y_i^*-Y_j^*\|,
\end{align*}
with $\{Y_1^*,\ldots,Y_n^*\}$ simulated from $\mathrm{P}_Y$. This estimated energy $\hat{E}^*$ can be compared with appropriate Monte Carlo simulation $\{Y_1^{*b},\ldots,Y_n^{*b}\}$, $b=1,\ldots,B$ designed to build an $\alpha$-level critical point using $\{\hat{E}^{*b}\}_{b=1}^B$. Note that, under the null hypothesis, $Y^*$ is simulated from $\mathrm{P}_{0}$. In the case of testing a composite hypothesis, then generation is done under $\mathrm{P}_{\hat\theta}$ with $\hat\theta$ estimated using $\{X_1,\ldots,X_n\}$.\\

Due to the scarcity of distribution models for random functions, the Gaussian case is one of the most widely studied, as it can bee seen, e.g., in \cite{Kellner19,Kol21} and in the recent review by \cite{Gore20} on tests for Gaussianity of functional data.\\

Finally, it is worth it to mention the two-sample problem, a common offspring of the simple-hypothesis one-sample GoF problem. Two-sample tests have also received a significant deal of attention in the last decades; see, e.g., the recent contributions by \cite{Jiang19} and \cite{Qiu21}, and references therein.

\section{GoF for regression models with functional data}
\label{GonzalezManteiga:3}

We assume henceforth, without loss of generality and for the sake of easier presentation, that both the predictor $X$ and response $Y$ are centered, so that the intercepts of the linear functional regression models are null.

\subsection{Scalar response}
\label{GonzalezManteiga:31}

A particular case of a regression model with functional predictor and scalar response is the so-called functional linear model. For $\mathcal{H}_X=L^2[0,1]$, this parametric model is given by
\begin{align}
	Y=m_\beta(X)+\varepsilon,\quad m_\beta(x)=\langle x,\beta\rangle=\int_0^1x(t)\beta(t)\,\mathrm{d}t, \label{eq:flm}
\end{align}
for some unknown $\beta\in\mathcal{H}_X$ indexing the functional form of the model. This popular model can be seen as the natural extension of the classical linear (Euclidean) regression model.\\

In general, there have been two mainstream approaches for performing inference on \eqref{eq:flm}: (\textit{i}) testing the significance of the trend within the linear model, i.e., testing $H_0:m\in\{m_{\beta_0}\}$ vs. $H_1:m\in\{m_\beta:\beta\in\mathcal{H}_X,\beta\neq\beta_0\}$, usually with $\beta_0=0$; (\textit{ii}) testing the linearity of $m$, i.e., testing $H_0:m\in\mathcal{L}=\{m_\beta:\beta\in\mathcal{H}_X\}$ vs. $H_1:m\not\in\mathcal{L}$.\\

For the GoF testing problem presented in (\textit{ii}), given an iid sample $\{(X_i,Y_i)\}_{i=1}^n$, one may consider the adaptation to this setting of the smoothing-based tests, with a basic test statistic structure given by $T_n=T(m_{nh},m_{\hat\beta})$, where $\hat\beta$ is a suitable estimator for $\beta$ and
\begin{align}
	m_{nh}(x)=\sum_{i=1}^nW_{ni}(x)Y_i=\sum_{i=1}^n\frac{K_h(\|x-X_i\|)}{\sum_{j=1}^nK_h(\|x-X_j\|)}Y_i \label{eq:NW}
\end{align}
is the Nadaraya--Watson estimator with a functional predictor. A particular smoothing-based test statistic is given by that of \cite{Delsol11},
\begin{align*}
	T_n=\int\left(m_{nh}(x)-m_{nh,\hat\beta}(x)\right)^2\omega(x)\,\mathrm{d}\mathrm{P}_X(x),
\end{align*}
which employs a weighted $L^2$ distance between \eqref{eq:NW} and $m_{nh,\hat\beta}$, the latter being a smoothed version of the parametric estimator that follows by replacing $Y_i$ with $m_{\hat\beta}(X_i)$ in \eqref{eq:NW}. Note that a crucial problem for implementing this test is the computation of the critical region $\{T_n>c_\alpha\}$, which depends on the selection of $h$ when a class of estimators for $\beta$ is used under the null. This class of smoothed-based tests were deeply studied in the Euclidean setting (see \cite{Gonzalez13}). Nevertheless, this is not the case in the functional context, except for the recent contributions by \cite{Maistre20} and \cite{Patilea20}.\\

As also presented by \cite{Gonzalez13} in their review, it is possible to avoid the bandwidth selection problem using tests based on empirical regression processes. For this purpose, a key element is the empirical counterpart of the integrated regression function $I_n(x)=n^{-1}\sum_{i=1}^n\mathbb{I}(X_i\leq x)Y_i$, where $X_i\leq x$ means that $X_i(t)\leq x(t)$, for all $t\in[0,1]$. In this scenario, the test statistic can be formulated as $T_n(I_n,I_{\hat\beta})$, where $I_{\hat\beta}(x)=n^{-1}\sum_{i=1}^n\mathbb{I}(X_i\leq x)\hat{Y}_i$, where $\hat{Y}_i=\langle X_i,\hat\beta\rangle$. Deriving the theoretical behavior of an empirical regression process indexed by $x\in \mathcal{H}_X$, namely $R_n(x)=\sqrt n(I_n(x)-I_{\hat\beta}(x))$ is a challenging task. Yet, as previously presented, the projection approach over $\mathcal{H}_X$ can be considered. The null hypothesis $H_0:m\in\mathcal{L}$ can be formulated as
\begin{align*}
	H_0:\mathbb{E}[(Y-\langle X,\beta\rangle)\mathbb{I}(\langle X,\gamma\rangle\leq u)]=0,\text{ for a $\beta\in\mathcal{H}_X$ and for all $\gamma\in \mathcal{H}_X$},
\end{align*}
which in turn is equivalent to replacing `for all $\gamma\in \mathcal{H}_X$' with `for all $\gamma\in \mathcal{S}_{\mathcal{H}_X}$' or `for all $\gamma\in \mathcal{S}_{\mathcal{H}_X,\{\psi_j\}_{j=1}^\infty}^{p-1},$ for all $p\geq1$', where
\begin{align*}
	\mathcal{S}_{\mathcal{H}_X}=\{\rho\in\mathcal{H}_X:\|\rho\|=1\},\quad
	\mathcal{S}_{\mathcal{H}_X,\{\psi_j\}_{j=1}^\infty}^{p-1}=\bigg\{\rho=\sum_{j=1}^p r_j\psi_j:\|\rho\|=1\bigg\}
\end{align*}
are infinite- and finite-dimensional spheres on $\mathcal{H}_X$, $\{\psi_j\}_{j=1}^\infty$ is an orthonormal basis for $\mathcal{H}_X$, and $\{r_j\}_{j=1}^p\subset\mathbb{R}$. As follows from \cite{Garcia14}, a general test statistic can be built aggregating all the projections within a certain subspace: $T_n=\int T_{n,\pi}\,\mathrm{d}W(\pi)$ with $T_{n,\pi}=T(I_{n,\pi},I_{\hat\beta,\pi})$ based on
\begin{align}
	\!\!\!I_{n,\pi}(u)=n^{-1}\sum_{i=1}^n\mathbb{I}(\pi(X_i)\leq u)Y_i\text{ and }
	I_{\hat\beta,\pi}(u)=n^{-1}\sum_{i=1}^n\mathbb{I}(\pi(X_i)\leq u)\hat{Y}_i, \label{eq:In}
\end{align}
for $\pi(x)=\langle x,\gamma\rangle$. In this case, $W$ is a probability measure defined in $\mathcal{S}_{\mathcal{H}_X}$ or $\mathcal{S}_{\mathcal{H}_X,\{\psi_j\}_{j=1}^\infty}^{p-1}$, for a certain $p\geq1$. Alternatively, the test statistic can be based on only one random projection: $T_n=T_{n,\hat\pi}$. More generally, $T_n$ may consider the aggregation of a finite number of random projections, as advocated in the test statistic of \cite{Cuesta19}. Both types of tests, all-projections and finite-random-projections, may feature several distances for $T$, such as Kolmogorov--Smirnov or Cramér--von Mises types.\\

Model \eqref{eq:flm} can be generalized to include a more flexible trend component, for instance, with an additive formulation. The functional generalized additive model (see \cite{McLean15}) is formulated~as
\begin{align}
	Y=m_F(X)+\varepsilon,\quad m_F(x)=\eta+\int_0^1F(X(t),t)\,\mathrm{d}t \label{eq:fgam}
\end{align}
and it can be seen that \eqref{eq:flm} is a particular case of \eqref{eq:fgam} with $F(x,t)=x\beta(t)$ and $\eta=0$. The functional $F$ can be approximated as
\begin{align*}
	F(x,t)=\sum_{j=1}^{k_X}\sum_{k=1}^{k_T}\theta_{jk}B_j^X(x)B_k^T(t),
\end{align*}
where $\theta_{jk}$ are unknown tensor product B-spline coefficients. Both for the $x$ and $t$ components, cubic B-spline bases, namely $\{B_j^X(x)\}_{j=1}^{k_X}$ and $\{B_k^T(t)\}_{k=1}^{k_T}$, are considered.\\

Model \eqref{eq:fgam} can be written in an approximated way as a linear model with random effects (see \cite{Tek19}) using the evaluations of $X_i(t_{in})$ over a grid $\{t_{in}\}\subset[0,1]$. Under the assumption of $\varepsilon$ being a Gaussian process, the so-called restricted likelihood ratio test (RLRT) can be used, where testing the GoF of the functional linear model \eqref{eq:flm} against model specifications within \eqref{eq:fgam} is equivalent to test that the variance of the random effect is null. \\

Another generalization of the functional linear model is given by the functional quadratic regression model introduced by \cite{Hor13}:
\begin{align}
	Y=\int_0^1\beta(t)X(t)\,\mathrm{d}t+\int_{0}^1\int_0^1\gamma(s,t)X(t)X(s)\,\mathrm{d}t\,\mathrm{d}s+\varepsilon. \label{eq:fq}
\end{align}
Clearly, when $\gamma=0$, \eqref{eq:flm} follows as a particular case of \eqref{eq:fq}. Using a principal component analysis methodology to approximate the covariance function $\mbox{Cov}(t,s)=\mathbb{E}[(X_i(t)-\mathbb{E}[X(t)])(X_i(s)-\mathbb{E}[X(s)])]$ with $\beta(t)=\sum_{j=1}^pb_jv_j(t)$ and $\gamma(s,t)=\sum_{j=1}^p\sum_{k=1}^pa_{jk}v_k(s)v_j(t)$ with $v_j$ the eigenfunctions of $\mbox{Cov}(t,s)$, model \eqref{eq:fq} can be written as a kind of linear model, were the null hypothesis $\gamma=0$ is tested.\\

A recent contribution by \cite{Lai20} is devoted to the testing a modified null hypothesis: $\widetilde H_0: ``X$ is independent of $\varepsilon$ and $m\in\mathcal{L}$'', using the recent results related with the distance covariance (see \cite{Sek07}, \cite{Lyons13}, and \cite{Sej13}). Consider $(\mathcal X,\rho_{\widetilde X})$ and $(\mathcal Y,\rho_{\widetilde Y})$ two semimetric spaces of negative type, where $\rho_{\widetilde X}$ and $\rho_{\widetilde Y}$ are the corresponding semimetrics. Denote by $(\widetilde X,\widetilde Y)$ a random element with joint distribution $\mathrm{P}_{\widetilde X\widetilde Y}$ and marginals $\mathrm{P}_{\widetilde X}$ and $\mathrm{P}_{\widetilde Y}$, respectively, and take $(\widetilde X',\widetilde Y')$ an iid copy of $(\widetilde X,\widetilde Y)$. The generalized distance covariance $(\widetilde X,\widetilde Y)$ is given by
\begin{align*}
	\theta(\widetilde X,\widetilde Y)=&\;\mathbb{E}\big[\rho_{\widetilde X}(\widetilde X, \widetilde X')\rho_{\widetilde Y}(\widetilde Y,\widetilde Y')\big]\\
	&+\mathbb{E}\big[\rho_{\widetilde X}(\widetilde X, \widetilde X')\big]\mathbb{E}\big[\rho_{\widetilde Y}(\widetilde Y,\widetilde Y')\big]\\
	&-2\mathbb{E}_{\widetilde X\widetilde Y}\big[\mathbb{E}_{\widetilde X'}\big[\rho_{\widetilde X}(\widetilde X,\widetilde X')\big]\mathbb{E}_{\widetilde Y'}\big[\rho_{\widetilde Y}(\widetilde Y,\widetilde Y')\big]\big].
\end{align*}

As noted by \cite{Lai20}, the generalized distance covariance can be alternatively written as
\begin{align*}
	\theta(\widetilde X,\widetilde Y)=\int\rho_{\widetilde X}(\widetilde x,\widetilde x')\rho_{\widetilde Y}(\widetilde y,\widetilde y')\,\mathrm{d}[(\mathrm{P}_{\widetilde X\widetilde Y}-\mathrm{P}_{\widetilde X}\mathrm{P}_{\widetilde Y})\times (\mathrm{P}_{\widetilde X \widetilde Y}-\mathrm{P}_{\widetilde  X}\mathrm{P}_{\widetilde Y})].
\end{align*}
Note that $\theta(\widetilde X,\widetilde Y)=0$ if and only if $\widetilde X$ and $\widetilde Y$ are independent. Given an iid sample $\{(\widetilde X_i,\widetilde Y_i)\}_{i=1}^n$ of $(\widetilde X,\widetilde Y)$, an empirical estimator of $\theta$ is given by
\begin{align*}
	\theta_n(\widetilde X,\widetilde Y)=\frac{1}{n^2}\sum_{i,j}k_{ij}\ell_{ij}+\frac{1}{n^4}\sum_{i,j,q,\tau}k_{ij}\ell_{q\tau}-\frac{2}{n^3}\sum_{i,j,q}k_{ij}\ell_{iq}
\end{align*}
with $k_{ij}=\rho_{\widetilde X}(\widetilde X_i,\widetilde X_j)$ and $\ell_{ij}=\rho_{\widetilde Y}(\widetilde Y_i,\widetilde Y_j)$. Taking $\widetilde X=X$ and $\widetilde Y=\varepsilon=Y-\langle X,\beta\rangle$, $\rho_{\widetilde Y}$ is the absolute value and $\rho_{\widetilde X}$ is the distance associated to $\mathcal{H}_X$. The test statistic is $T_n=\theta_n(\hat\varepsilon,X)$ and is based on $\{(X_i,Y_i-\langle X_i,\hat\beta\rangle)\}_{i=1}^n$. \\

All the tests described in this section have challenging limit distributions and need to be calibrated with resampling techniques.

\subsection{Functional response}
\label{GonzalezManteiga:32}

When both the predictor and the response, $X$ and $Y$, are functional random variables evaluated in $\mathcal{H}_X=L^2[a,b]$ and $\mathcal{H}_Y=L^2[c,d]$, the regression model $Y=m(X)+\varepsilon$ is related with the operator $m:\mathcal{H}_X\rightarrow \mathcal{H}_Y$. Perhaps the most popular operator specification is a (linear) Hilbert--Schmidt integral operator, expressible as 
\begin{align}
	m_\beta(x)(t)=\langle x,\beta(\cdot,t)\rangle=\int_a^b\beta(s,t)x(s)\,\mathrm{d}s,\quad t\in[c,d], \label{eq:flmf}
\end{align}
for $\beta\in\mathcal{H}_X\otimes\mathcal{H}_Y$, which is simply referred to as the functional linear model with functional response. The kernel $\beta$ can be represented as $\beta = \sum_{j=1}^\infty \sum_{k=1}^\infty b_{jk} (\psi_j \otimes \phi_k)$, with $\{\psi_j\}_{j=1}^\infty$ and $\{\phi_k\}_{k=1}^\infty$ being orthonormal bases of $\mathcal{H}_X$ and $\mathcal{H}_Y$, respectively.\\

Similarly to the case with scalar response, performing inference on \eqref{eq:flmf} have attracted the analogous two mainstream approaches: (\textit{i}) testing $H_0:m\in\{m_{\beta_0}\}$ vs. $H_1:m\in\{m_\beta:\beta\in\mathcal{H}_X\otimes\mathcal{H}_Y,\beta\neq\beta_0\}$, usually with $\beta_0=0$;  (\textit{ii}) testing $H_0:m\in\mathcal{L}=\{m_\beta:\beta\in\mathcal{H}_X\otimes\mathcal{H}_Y\}$ vs. $H_1:m\not\in\mathcal{L}$. The GoF problem given in (\textit{ii}) can be approached by considering a double-projection mechanism based on $\pi_X:\mathcal{H}_X\rightarrow\mathbb{R}$ and $\pi_Y:\mathcal{H}_Y\rightarrow\mathbb{R}$. Given an iid sample $\{(X_i,Y_i)\}_{i=1}^n$, a general test statistic follows (see \cite{Garcia20}) as $T_n=\int T_{n,\pi_X,\pi_Y}\,\mathrm{d}W(\pi_X\times\pi_Y)$ with $T_{n,\pi_X,\pi_Y}=T(I_{n,\pi_X,\pi_Y},I_{\hat\beta,\pi_X,\pi_Y})$, where $I_{n,\pi_1,\pi_2}$ and $I_{\hat\beta,\pi_1,\pi_2}$ follows from \eqref{eq:In} by replacing $\pi$ with $\pi_X$, and $Y_i$ and $\hat{Y}_i$ with $\pi_Y(Y_i)$ and $\pi_Y(\hat{Y}_i)$, respectively. In this case, $W$ is a probability measure is defined in $\mathcal{S}_{\mathcal{H}_X}\times \mathcal{S}_{\mathcal{H}_Y}$ or $\mathcal{S}_{\mathcal{H}_X,\{\psi_j\}_{j=1}^\infty}^{p-1}\times\mathcal{S}_{\mathcal{H}_Y,\{\phi_k\}_{k=1}^\infty}^{q-1}$, for certain $p,q\geq1$. The projection approach is immediately adaptable to the GoF of \eqref{eq:flmf} with $\mathcal{H}_X=\mathbb{R}$, and allows graphical tools for that can help detecting the deviations from the null, see \cite{Garcia20a}. An alternative route\nopagebreak[4] considering projections just for $X$ is presented by \cite{Chen20}.\\

The above generalization to the case of functional response is certainly more difficult for the class of tests based on the likelihood ratios. Regarding the smoothing-based tests, \cite{Patilea16} introduced a kernel-based significance test consistent for nonlinear alternative. More recently, \cite{Lee20} proposed a significance test based on correlation distance ideas.

\section*{Acknowledgements}

The authors acknowledge the support of project MTM2016-76969-P, PGC2018-097284-B-100, and IJCI-2017-32005 from the Spain's Ministry of Economy and Competitiveness. All three grants were partially co-funded by the European Regional Development Fund (ERDF). The support by Competitive Reference Groups 2017--2020 (ED431C 2017/38) from the Xunta de Galicia through the ERDF is also acknowledged.


\end{document}